\documentclass[namedreferences]{SolarPhysics}
\usepackage{lscape,epsfig}
\usepackage{url}
\usepackage[optionalrh]{spr-sola-addons}
\begin{document}
\begin{article}
\begin{opening}

\title{Web-Based Data Processing System for Automated Detection of Oscillations with Applications to the Solar Atmosphere}

\author{R.A. \surname{Sych}$^{1}$\sep
        V.M. \surname{Nakariakov}$^{2}$\sep
        S. \surname{Anfinogentov}$^1$\sep
        L. \surname{Ofman}$^3$
        }
\runningauthor{R.A. Sych \textit{et al.}} \runningtitle{Web-Based Data Processing System}

\institute{$^{1}$  Institute of Solar-Terrestrial Physics, 126,
Lermontov St., Irkutsk, Russia,
           email: \url{sych@iszf.irk.ru}\\
           $^{2}$ Physics Department, University of Warwick, Coventry CV4 7AL,UK,
           email: \url{V.Nakariakov@warwick.ac.uk}\\
           $^{3}$ The Catholic University and NASA Goddard Space Flight Center, Greenbelt, MD 20771, USA;
           email: \url{Leon.Ofman@nasa.gov}\\
                       }

\date{Received .. ; accepted .. }

\begin{abstract}
A web-based, interactive system for the remote processing of imaging
data sets (\textit{i.e.}, EUV, X-ray and microwave) and the automated
interactive detection of wave and oscillatory phenomena in the solar
atmosphere is presented. The system targets localised, but spatially
resolved, phenomena, such as kink, sausage, and longitudinal
propagating and standing waves. The system implements the methods of
Periodmapping for pre-analysis, and Pixelised Wavelet Filtering for
detailed analysis of the imaging data cubes. The system is
implemented on the dedicated data processing server
\url{http://pwf.iszf.irk.ru}, which is situated at the Institute of
Solar-Terrestrial Physics, Irkutsk, Russia. The input data in the
\textsf{.sav}, \textsf{.fits} or \textsf{.txt} formats can be submitted via the local and/or
global network (the Internet). The output data can be in the \textsf{png},
\textsf{jpeg} and binary formats, on the user's request. The output data are
periodmaps; narrowband amplitude, power, phase and correlation maps
of the wave's sources at significant harmonics and in the chosen
spectral intervals, and \textsf{mpeg}-movies of their evolution.  The system
was tested by the analysis throughout the EUV and microwave emission from
the active region NOAA 10756 on 4 May 2005 observed with TRACE and
the Nobeyama Radioheliograph. The similarity of the spatial
localisation of three-minute propagating waves, near the footpoint of
locally open magnetic field lines determined by the potential field
extrapolation, in both the transition region and the corona was
established. In the transition region the growth of the three-minute
amplitude was found to be accompanied by the decrease in the line of
sight angle to the wave propagation direction.
\end{abstract}

\end{opening}

\section{Introduction}
\label{Introduction} The detection and the classification of various
solar features is a complex task that relies on the expertise of
observers, modelers, and theorists. The identification of highly
energetic features, such as flares and CMEs, is often
straightforward, and their signatures are well known and usually
uniformly accepted in the solar physics community. Also, large-scale
features, such as coronal holes, active regions, and loops can be
identified with standard image processing algorithms (\textit{e.g.} Aschwanden \textit{et
al.}, 2008; Scholl and Habbal, 2008; Krista and Gallagher, 2009).
The main difficulty lies in the identification of \lq\lq
weak" features, which are close to the instrumental resolution or
signal-to-noise threshold. However, these features, such as waves,
often can provide us with invaluable information on the physical
conditions in the solar atmosphere (\textit{e.g.} see the review by Nakariakov and
Verwichte, 2005).

The study of coronal waves and oscillations is one of the most
rapidly developing areas of solar physics and lies in the mainstream
of the scientific exploitation of solar mission data.  Coronal-wave
processes attract our attention because of their possible role in
the enigmatic problems of coronal physics: coronal heating,
triggering of flares and CMEs, and the acceleration of the solar
wind. Also, as properties of waves contain information about the
medium of their propagation, the waves are natural probes of the
medium, providing us with a tool for its diagnostics. The analysis
of coronal data in order to determine parameters of waves and
oscillations is a non-trivial task, mainly because of the detection
of these phenomena near the very threshold of the instrumental
detectability, and the necessity to analyse the spatial, temporal,
and phase variations simultaneously at the highest-possible
resolution. The problem has been amplified by the lack of an
instrument dedicated to coronal-wave studies. The \textit{Solar Dynamics
Observatory/Atmospheric Imaging Assembly} (SDO/AIA) is the first
instrument to have coronal seismology as one of its scientific
themes. AIA will provide us with greatly enhanced cadence and
signal-to-noise ratio in comparison with the previous generation of
solar-coronal imagers. In the past, the detection of waves in the
solar corona relied heavily on individual examination of time
sequences of images (\lq\lq movies") by experts in the field. The
availability of AIA will substantially increase the data rates,
requiring the development of novel data analysis methods for the
automated detection of coronal wave and oscillatory
processes. The pioneering studies of this problem
(De Moortel and McAteer, 2004; Nakariakov and King, 2007; Marsh \textit{et al.}, 2008; McIntosh \textit{et al.}, 2008) have addressed this issue,
investigating automated-detection techniques of coronal waves, based upon the wavelet and Fourier transforms, Bayesian statistics and
coherence/travel time based approach.

There are several classes of MHD waves in the solar corona, which
are to be recognized by an automated detection system. The
observational evidence of the presence of MHD waves and oscillations
is abundant (see \textit{e.g.} Nakariakov and Verwichte, 2005). Observed
properties of these waves are determined by the original MHD mode
(slow or fast magnetoacoustic, Alfv\'en) and the guiding plasma
structure. Different MHD modes of plasma structures have very
different physical properties (\textit{e.g.} compressibility, polarization,
dispersion, cut-offs, \textit{etc.}) and hence have different observational
manifestations.  SDO/AIA will be able to see at least two of them,
the transverse (kink) and the longitudinal modes at higher spatial
and temporal resolution than previously.

In this paper we describe a web-based system for the remote processing of imaging datasets (such as SOHO/EIT, TRACE, STEREO/EUVI, Nobeyama Radioheliograph, \textit{Hinode}/XRT, CORONAS-PHOTON/TESIS, and SDO/AIA, designed for the automated detection of periodic and quasi-periodic processes. The system has a modular structure that allows for the easy update and inclusion of new data analysis and visualization tools. The present version of the system implements the techniques of periodmapping (Nakariakov and King, 2007) and the Pixelised Wavelet Filtering (PWF: Sych and Nakariakov, 2008). Also, there is a possibility of using the standard Morlet wavelet and periodogram analysis of 1D signals.

The method of peridomapping provides fast preliminary
analysis of imaging datacubes, allowing for the identification of
the candidate regions and time intervals for the presence of
spatially resolved oscillatory phenomena. With this method, the time
is divided into 30 minutes (or, optionally, of other duration) time
slots which overlap by a half of the duration. Periodograms of the
time signal of each pixel (or optionally a macropixel, consisting of
several adjacent pixels for improved signal-to-noise ratio) are
constructed. Options include signal smoothing, user-defined
filtering, and construction of the running-difference signal. Then,
the frequency of the highest spectral power is identified and tested
against a certain criterion. The default  is that if the
highest spectral power is higher than five times the mean level of the
spectral power, then the signal is treated as an \lq\lq oscillatory"
and the frequency of the highest spectral power is determined. Then,
the pixel is assigned  a certain colour according to a chosen
colour coding scheme, in which different colours correspond to
different spectral frequencies. If the highest spectral power is lower than the detection threshold, the signal is treated as \lq\lq
non-oscillatory" and the pixel is left blank. Hence, a static 2D map
of the analysed field of view (a \lq\lq periodmap") for the analysed
time interval is created. The blobs of the same or similar colour
(corresponding to the same or similar time frequency of the
oscillation) indicate the presence of the oscillation. Then the
process is repeated for the next time interval. As a result, a time
sequence of periodmaps is obtained, which can be very easily
analysed by visual inspection, or automatically,
identifying the \lq\lq oscillatory regions" in certain time
intervals.

The PWF method (Sych and Nakariakov, 2008) is a generalisation of
the wavelet transform of 3D datacubes. Temporal signal of each spatial
pixel is wavelet transformed (\textit{e.g.} with a Morlet mother function),
which results in the power, amplitude, and phase 4D data cubes (two
spatial dimensions, time, and the frequency). The 4D data cubes can
be processed according to a specific request. For example,
selecting a certain spectral component or integrating over a certain
narrow spectral range makes a 3D narrowband datacube that consists
of a sequence of narrowband maps. The temporal evolution of the sources
of the narrowband oscillations can be visualised by a movie (\textit{e.g.} in
the MPEG format). Similarly, one can create a movie of the global
wavelet spectrum evolution. Selecting a certain 1D slit on the map
makes it possible to make narrowband or broadband time--distance
plots. In particular, this allows one to identify the observed mode,
to discriminate between standing and propagating modes, determine
the morphology of wave sources and study its time evolution, and
detect multiple periodicities and their interrelation. This
knowledge is crucial for MHD coronal seismology and the study of the
solar atmospheric connectivity. In the future, we plan to include
other data processing tools, such as the Gradient Pattern Analysis
(GPA-method: Rosa \textit{et al.}, 2000).

The main aim of the present system is the remote processing of
data cubes supplied by the user.  The user does not need to have
deep knowledge of data processing techniques, and programming. Also,
it is not necessary to have the Interactive Data Language (IDL)
installed on the user's computer. The minimal prerequisites are
access to the Internet and a standard web browser. The input
parameters are the imaging data cubes. The output parameters are the
global and local (\textit{e.g.} for a selected pixel or macropixel, or a
slit) spectra, narrowband images, periodmaps, wave vector maps, and
movies showing the temporal evolution of the maps.

Application of the system to the detection of coronal waves and
oscillations is illustrated with the use of the microwave and EUV
imaging data cubes of NOAA 10756 obtained with TRACE and the
Nobeyama Radioheliograph (NoRH). During the passage of this active
region across the solar disk on 1\,--\,5 May 2005 both standing and
propagating waves have been detected (see Sych \textit{et al.}, 2009 for
details), which made it a good test case for the present study.

The paper is organised as follows. In the next section we describe some existing examples of on-line data access and analysis tools.
Section 3 describes the scientific objectives of the system. Section 4 discusses the system architecture and gives an example of the user
interface. Sections 5 and 6 illustrate the application of the system to the analysis of observational data from the spatial and temporal points of view, respectively. The conclusions and discussion of the results obtained are given in Section 7.

\section{Existing On-Line Resources}

In the design of the present system, existing on-line systems for
information processing were used. At the moment, on the Internet,
there are a number of sites for the remote processing of documents,
multi-lingual translators, search engines, on-line game resources
etc. Examples of such sites are the following.

\url{http://www.aie.sp.ru/Calculator_filter.html} - automatic calculation of the filtering elements for radio-engineering systems. The output is the amplitude and phase spectral properties of the designed system, and the integral properties of the circuit.

\url{http://ion.researchsystems.com/IONScript/wavelet}  - on-line interactive continuous wavelet transform for time-series analysis.

\url{http://translate.google.com/translate}  - multi-lingual translation of web resources

\url{http://docs.google.com} - Create and share work online, edit document any time, from anywhere.  Share changes in real time.   Files are stored securely online.

In the solar research community, examples of such resources are  the
Virtual Solar Observatory [\url{http://umbra.nscom.nasa.gov/vso/}]
that provides one with the access to the data of different solar
observational facilities; Space Weather Monitoring
[\url{http://sidc.oma.be/html/LatestSWData.html}] that provides real-time information on the solar, heliospheric and magnetospheric
activity and Solar Monitor [\url{http://solarmonitor.org/index.php}] -
that provides the monitoring of active regions in the solar
atmosphere. These resources aim at the data warehousing and data
search and access optimisation, and usually do not process the data.
Thus, our system is different from the existing on-line
tools.

\section{The Proposed Scheme of Automated Detection of Oscillatory Features in Datacubes }

The combination of the periodmapping and PWF techniques suggests the following method for the automated detection of coronal waves and oscillations in imaging data cubes:

\begin{figure}    %%%%%%%%%%%%%%%%%% FIGURE 1
%\hbox{ \includegraphics[viewport=0 206 554 793] \psfig{file=fig1.eps,width=12cm,clip=}}
\resizebox{\hsize}{!}{\includegraphics[viewport=0 206 554 793]{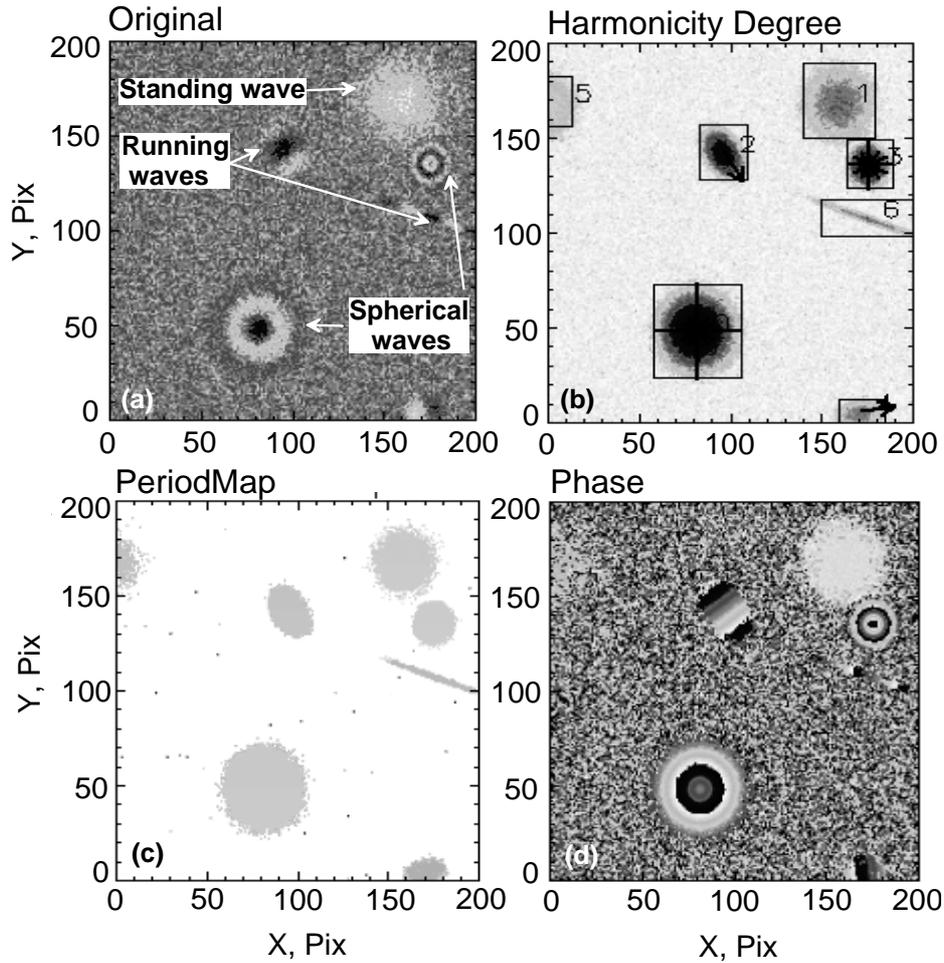}}
\caption{Application of the wave-activity recognition method. Panel (a) shows a
snapshot of an artificial test data cube with standing, and
spherical and plane running waves. Panels (b) and (c) shows the
harmonicity degree and the periodmap of the data cube, respectively.
The vectors show the local direction of the wave propagation.  Panel
(d) shows a snapshot of the phase distribution.}
\label{fig1}
\end{figure}

\begin{itemize}
\item[\textit{i)}] Full-disk data cubes of 30 and 60-minute duration,
overlapping in the middle of the sampling time interval, will be
processed with the Periodmapping technique. The outcome is a
sequence of static full-disk images which highlight the presence of
spatial regions and time intervals identified for detailed study,
aiming at detection of transverse and longitudinal oscillations and
waves. The periodmaps are then stored for further analysis. The
selection of the regions of interest (ROI) on the map will be
performed manually or with the implementation of an automated
pattern-recognition method. The comparison of possible techniques
can be found in Aschwanden \textit{et al.} (2008).

\item[\textit{ii)}] The specific pattern-recognition technique built in the
system is based upon the application of the harmonicity
criterion.  For this we define minimum $\epsilon_\mathrm{min}$, maximum $\epsilon_\mathrm{max}$ and average
$\epsilon_\mathrm{ave}$ values of the peaks in Fourier power spectra of each pixel
and then calculate the harmonicity degree $H \equiv (\epsilon_\mathrm{max}-\epsilon_\mathrm{ave})/(\epsilon_\mathrm{ave}-\epsilon_\mathrm{min})$.
The spatial distribution of these values forms a harmonicity
degree map of the datacube, highlighting  the regions of high harmonicity. Then, we determine spatial regions
where the harmonicity degree is higher than a certain threshold value. Regions smaller
than some prescribed value can be disregarded. Then we may inscribe the regions of the
enhanced harmonicity into circles or rectangles to get an ROI of a simple shape.
From the Fourier transform
of the signal of each pixel of the ROI we calculate the phase
$\Theta\equiv \Im(\ln(z))$, where $z$ is the complex amplitude of the highest amplitude spectral component.
Also, it is possible to calculate the phase differences between the neighbouring pixels in a certain prescribed direction
(\textit{e.g.} horizontal or vertical), constructing phase-difference maps.
On such a map, a standing-wave pattern corresponds to
the phase vector of zero magnitude, in contrast to propagating
waves which have nonzero values of the projected wave vectors. Also, for a running wave (with a significant
component of the wave vector in the plane perpendicular to the
line-of-sight (LOS)) one half of the wave source has positive values
of the phase difference, while it is negative in the other half.
This property allows for the automatic discrimination between the
standing, spherical, and plane running wave motions by constructing
the maps of the projected wave vectors.

The system was tested by detecting three different types of waves: plane running, cylindrical running or spherical and standing. Realistic conditions were reproduced by randomization of the parameters of the waves (the type of the wave, amplitude, size of the source, direction of propagation, times of appearance and disappearance), as well as by including noise and time modulation.  Figure 1 shows a snapshot of the system outcome at a certain instant of time. The types of recognised waves are labelled. In each test, a new built-in wave pattern was generated.

All wave patterns are spatially localised. Periodmapping and Harmonicity Degree maps of the data cube
(Figure~\ref{fig1}c) and (Figure~\ref{fig1}b) identify the
location of the wave patters. A snapshot of the phase
distribution is shown in Figure~\ref{fig1}d.

\item[\textit{iii)}] The comprehensive analysis of selected ROI is carried out
with the use of the PWF method, which allows for the detailed study
of all aspects of the wave's evolution and interaction.
\end{itemize}

\section{The System Architecture and User Interface}

The key principles of the system architecture are that all
calculations should be carried out directly on the server remotely
from the user through the Internet; the software and hardware
prerequisites for the user's computer should be minimal; the user
should be able to upload and download data; the calculations on the
server are carried out with the use of the \emph{Interactive Data
Language} (IDL). Hence it was decided to design the system as a
web application, with the use of a standard Internet browser as the
user's client. From the user's point of view, the system looks like
an Internet site. The system is controlled (including data uploading
and downloading) from the web page.

%The performance assessment of a computational system is based upon the dependence of the quality and authenticity of results obtained on the input signal. In the considered case, it is the data cubes prepared by the users. The preparation should include de-trending and de-rotation, accounting for known instrumental artefacts (e.g. the non-uniformity of the sensitivity, dark current, antenna sidelobes), which can corrupt the system performance.
The presented system is made more user-friendly
by reducing the effect of some known mistakes connected with poor preparation of the input data
or with the lack of user experience, which may lead to the generation of artifacts in the output.
In particular, the number of possible mother wavelet functions is limited; an optimal
number of wave numbers is set up; the variation of the frequency resolution is restricted.
In any case, the final interpretation of the results obtained with the use of the system is entirely
up to the user.

\begin{figure}    %%%%%%%%%%%%%%%%%% FIGURE 2
\resizebox{\hsize}{!}{\includegraphics[viewport=56 101 527 730]{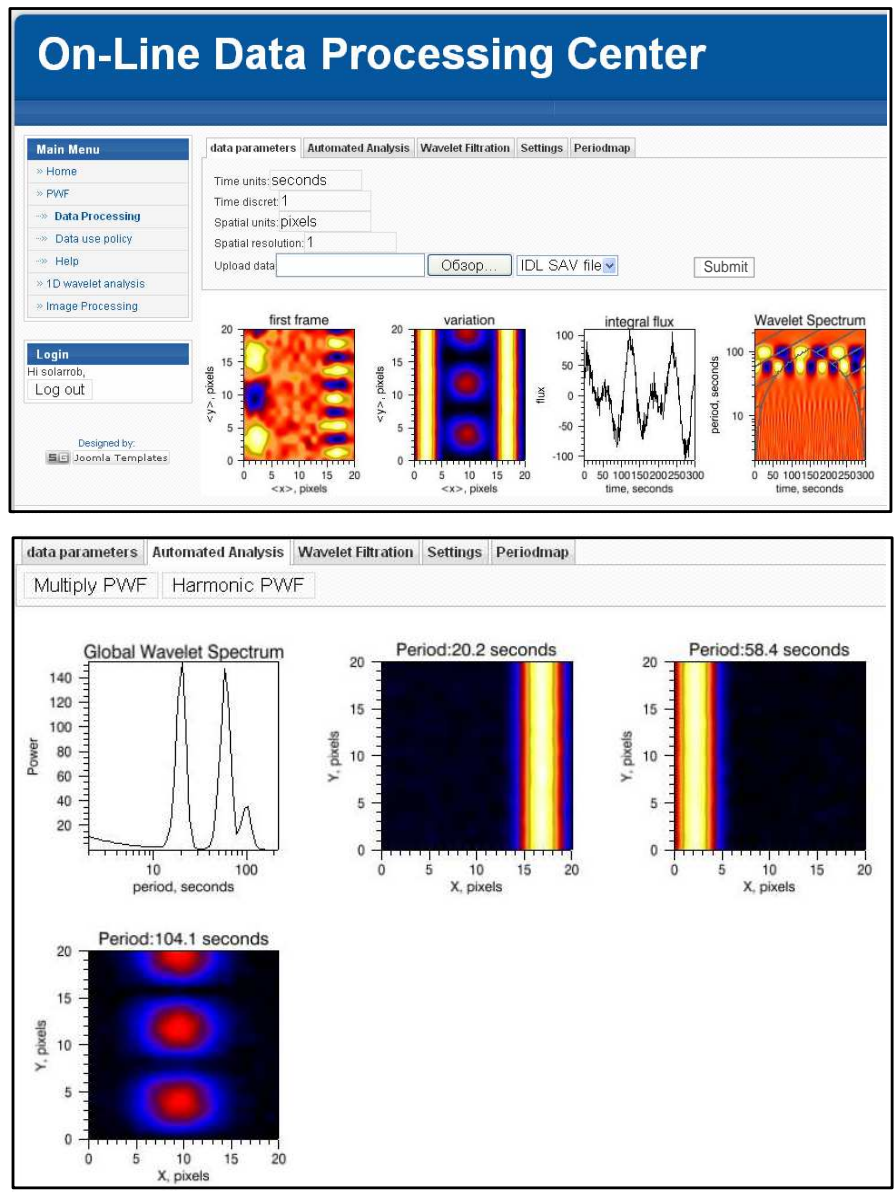}}
\caption{Upper panel: The main window of the On-Line Data Processing Centre. Bottom panel: The window of harmonic PWF data processing.}
\label{fig2}
\end{figure}

The key elements of the system are:
\begin{itemize}
  \item Internet browser for the user interface;
  \item Web server receiving the user's queries and transferring the input and output data;
  \item Server scripts responsible for the interaction between the web-server and the IDL codes which carry out the calculations;
  \item The codes that implement the data analysis and output generation techniques.
\end{itemize}

Taking into account these architecture principles, we created a web-site [\url{http://pwf.iszf.irk.ru}] that provides access to the remote data analysis server (Figure~\ref{fig2}). The specific implementation includes the following components:
\begin{itemize}
  \item Web-server {\sf Apache};
\item The web site is controlled with the {\sf CMS Joomla} software with an integrated component written in PHP, which carries out the interface with the IDL programmes;
\item IDL is used for that data processing, analysis and creation of the graphical and video files.
\end{itemize}

Step-by-step work of the system is as follows:
\begin{itemize}
\item   On the user's command, a temporary file folder for the storage of files (images and videos, temporary files, postscripts) of the current session is automatically created by a {\sf PHP} script and assigned with a unique name; the data are uploaded to the folder;
\item   Simultaneously, a new server session is started. The session data are stored in a {\sf MySQL} database. Session identification cookies are sent to the user's browser;
\item   After that, the {\sf PHP} script opens the {\sf IDL} interpreter and runs the required {\sf IDL} programmes, supplying them with the parameters according to the user's queries;
\item   The {\sf IDL} programme  reads the uploaded data, processes them, saves them in a file, and passes the control to the {\sf PHP} script;
\item   PHP script reads the output data from the hard drive and send it to the user's browser;
\item   At the end of the session, the outcome data are sent to the user, the temporary folder is stored for three days and then deleted.
\end{itemize}
The session-based approach allows for  simultaneous work by several users. Input, output, and temporary files of different users are stored in different folders.

Each data-analysis technique (wavelet, periodmapping, PWF) has a specific page on the web-site, which contains its description, user's manuals, references to the journal papers, and the description of the data analysis server interface. The interface allows the user to upload the data, and run the data analysis procedures.

The system performance was tested by processing time sequences of 512$\times$512 pixel images on a standard PC (Intel Core Quad Q9400 2.66 GHz, 4Gb RAM). The temporal window was 100 time steps, and the analysis was carried out for each ten time steps. The total speed of the processing, including the system initialization time and the file exchanges, was found to be 100 frames per minute, or 16 Mbit s$^{-1}$. The use of GPU tools will increase the total processing speed by 10--20 times. Thus, even a standard PC can be used for the high spatial and temporal resolution data processing.

Consider the work of the system using the PWF method as an example. To start, the user selects the item \emph{Data Processing} in the Main menu (Figure~\ref{fig2}), which opens the Data Processing screen. In the upper part of the screen, there is a control panel that contains tabs of different functions. In the main part of the screen the current data processing results are shown. The work begins with the use of the  \emph{Data Parameters} tab that allows the user to specify the temporal and spatial resolution of the input data cube, the units, and the format of the uploaded data. At the moment, the system supports three input data formats: ASCII, FITS, and SAVE (the internal IDL format). Also, here the user sees the form for the data uploading. Once the data are uploaded, the main screen shows the first image of the data cube, its variation map, the time dependence of the integrated flux (the \lq\lq light curve") and the amplitude global wavelet spectrum. The spectrum is constructed by the summation of the spectra of individual pixels. Such a spectrum can be quite different from the spectrum calculated for the integrated light curve that is obtained by the summation of time signals of the individual pixels. The reason for the discrepancy is that in the latter case the spectrum can be affected by the phase differences between the signals of different pixels, cancelling the signals.

The next tab, \emph{Automatic analysis}, contains the buttons for the run of two different algorithms of the PWF analysis, multi-component and harmonic. The multi-component PWF analysis is based upon the expansion of the temporal signal of each pixel by the octaves of the wavelet spectrum with the frequencies that are multiples of two. The number and the spectral width of the channels are determined by the cadence and the duration of the signal. For each frequency band, the spectral power of the signal is calculated. Repeating this for each pixel, we can synthesise a broadband oscillation map of the data cube. This approach allows for the rapid determination of the spatial-frequency distribution of the harmonic oscillation power in the signal, and its spatial localization. The second algorithm, the harmonic PWF, is based upon the construction of narrowband images of the oscillatory sources that correspond to certain selected harmonics (\textit{e.g.} corresponding to significant peaks in the global wavelet spectrum). In comparison with Fourier spectrum, the global wavelet spectra are known to be smoother in the high-frequency part of the spectrum, and better resolution in the lower-frequency part. Hence, the determination of the significant spectral peaks (\textit{e.g.} over the 90\% significance calculated according to Torrence and Compo, 1998), the use of global wavelet spectra reduces the level of the spurious information that is not connected with the significant harmonics.

The  flexibility allowed in the choice of the wavelet analysis parameters is based
upon the minimisation of their effect, allowing effective data processing by an
inexperienced user.  In the present version of the system, the user can choose the band
of analysed frequencies and frequency resolution and select certain harmonics of interest. The noise can be removed in two ways:  running averaging and median averaging in both  spatial and temporal domains. Images with a prescribed level of significance are obtained by the adjustment of the significance levels of each pixel to the prescribed value. In the images the pixels where the cutoff frequency, determined by the cone of influence, is  exceeded are highlighted \textit{e.g.} by framing.

Detailed analysis of the input data can be carried out in the manual
regime, which can be reached by the \emph{Wavelet Subband
Filtration} tab. Here the user can prescribe the band of the periods of
interest, as well as the temporal interval of interest. This approach
allows the user to study the amplitude and period modulation of
coronal waves and oscillations, minimizing the effect of
neighbouring frequencies, giving the time evolution of the
user-selected narrowband amplitude, power and phase maps in the form
of {\sf MPEG}-movies.

A periodmap of the analysed data cube is constructed in the \emph{Periodmap} tab, which allows the user to specify the upper and lower cut-off frequencies. General settings of the data analysis (\textit{e.g.} the filtering methods and thresholds, the possibility of subtracting the running mean) and of the visualization (the colour table, the format of the output file, \textit{e.g.} PostScript) can be modified on the  \emph{Settings} tab.

\section{Test Data (Spatial Approach)}

The system has been tested by the analysis of the data obtained with
TRACE in the 171\AA\ bandpass with the temporal resolution of eight seconds, and spatial resolution 1$^{\prime\prime}$. The test data sequence contained
the active region NOAA 10756 observed from 1 to 4 May 2005 during
its passage across the solar disk. The detailed analysis of this AR
is presented in Sych \textit{et al.} (2009). During the observation interval,
the AR showed high radio-burst activity, observed by NoRH in 17 GHz.
A number of radio bursts with duration from 10 seconds to 30 minutes were
detected. Also, stationary oscillations of the integrated radio flux
with the period about three minutes were detected. The three-minute oscillations
were found to be caused by the variations of the radio brightness of
the sunspot associated with the AR. The analysis revealed a
relationship between the sunspot oscillations and flaring energy
releases, in particular, radio bursts, in this AR and nearby.  Radio
bursts were found to be preceeded by the increase of the power of the
three-minute oscillations. A significant spectral peak was typically
appearing about 10\,--\,12 minutes before the burst. The PWF analysis of the
17 GHz NoRH datacube revealed that the increase in the three-minute
oscillation power was accompanied by the appearance of a
characteristic V-shaped structure on the three-minute narrowband maps. The
fingers of the structure were directed towards the burst location.
The V-shaped structures were found to coincide with the low-lying
magnetic loops linking the regions of the opposite magnetic
polarity. This finding suggests that the flaring energy releases
could be associated with longitudinal waves leaking from the sunspot
along the magnetic flux tubes towards the flaring site. Also, the
presence of a significant three-minute spectral component in the flaring
light curves indicates that the three-minute longitudinal waves can
modulate the emission.

Here we continue the study begun in Sych \textit{et al.} (2009), concentrating on the three-minute waves in quiet-Sun conditions. Initially, the TRACE 171\AA\ datacube was preprocessed with the use of the {\sf trace\_prep} subroutine (signal calibration, dark current/pedestal removal, CCD gain and lumogen correction, \textit{etc.}). Also, the data were derotated. Then, the datacube was analysed with the use of the system described.

\subsection{Periodmap Analysis of the Active Region NOAA 10756}

\begin{figure}    %%%%%%%%%%%%%%%%%% FIGURE 3
\resizebox{\hsize}{!}{\includegraphics[viewport=0 223 545 764]{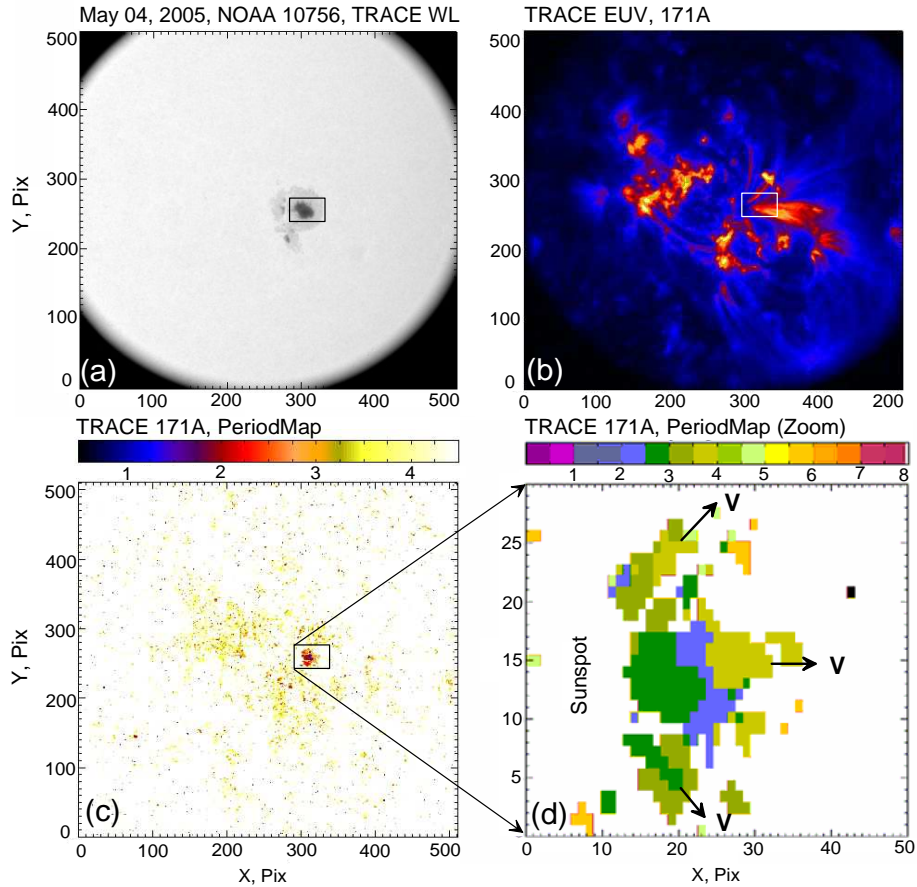}}
\caption{Top panels: (a) the active region NOAA 10756 in white light; (b) EUV 171\AA\ image of the ROI obtained with TRACE on 04 May 2005, 03:02 UT.  The square in panel (a) shows the position of the ROI (50$^{\prime\prime}\times$30$^{\prime\prime}$). Bottom panels:  (c) The periodmap of the EUV datacube at 03:30\,--\,03:50 UT; (d) enlarging the oscillating region. The horizontal bars represent the period scale in minutes. } \label{fig3}
\end{figure}

The preliminary analysis was carried out by the Periodmap technique.
Figure~\ref{fig3} shows the images of AR NOAA 10756 on 4 May 2005
taken at 03:02 UT in white light and in the EUV. It is evident that
in the corona the sunspot coincides with a system of bright
loops originating from its umbra. The periodmap of the TRACE 171\AA\
datacube at 03:30\,--\,03:50 UT indicates the presence of three-minute
oscillations situated over the sunspot, in the region highlighted by
the square. In the same region, the visual inspection of the
datacube movie reveals the presence of longitudinal waves
propagating outwards from the sunspot along the loops. Outside the
sunspot, there are small regions of significant oscillations, which
correspond to the footpoints of the loops in the active region.

The size of the spatial localisation of the oscillating regions
(\lq\lq \emph{wave traces}") is found to be about 20\,--\,30$^{\prime\prime}$, which
coincided with the results obtained by other authors (Berghmans and
Clette, 1999; De Moortel \textit{et al.}, 2000, 2002a,b;
Robbrecht \textit{et al.}, 2001; King \textit{et al.}, 2003; De Moortel, 2006). Also, in
agreement with previous studies of this phenomenon, the wave traces
are situated at the footpoint of the magnetic fan structure. Some
non-uniformity in the spatial distribution of the dominating periods
is revealed as well: shorter period oscillations (two\,--\,three minutes) are
apparently situated closer to the footpoint.

\subsection{PWF Analysis of the Active Region NOAA 10756}

\begin{figure}    %%%%%%%%%%%%%%%%%% FIGURE 4
\resizebox{\hsize}{!}{\includegraphics[viewport=0 175 554 775]{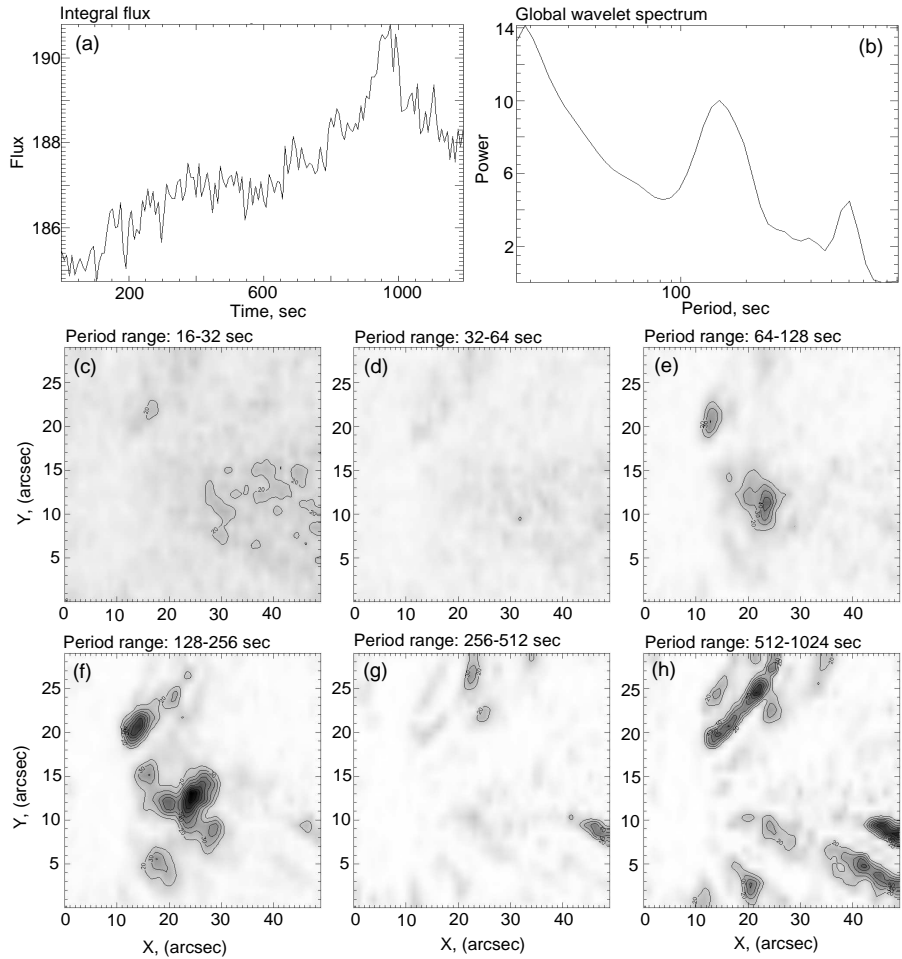}}
\caption{Examples of the PWF analysis of EUV TRACE 171\AA\ temporal data cube for the 04 May 2005 03:30\,--\,03:50 UT. The global wavelet spectrum of the time profile of integral flux (the light curve, (a)) is shown in (b). The panels (c\,--\,e) presents the sequence of narrowband maps constructed for six different ranges of periods.  The solid black contours  outline the regions with the maximal power of harmonic oscillations. The level contours represent the percentage of the maximum of the wideband spectral power. }
\label{fig4}
\end{figure}

The light curve of the total flux in the interval 03:30\,--\,03:50 UT
and its global wavelet spectrum (Figure~\ref{fig4}) demonstrate the
presence of three- and ten-minute oscillatory patterns.  More detailed
information is obtained with the use of the PWF technique.
The narrowband maps made with PWF in the six (multiple of two, from 16 seconds
to 1024~seconds) period bands, are shown in Figure~\ref{fig4}.

The global normalisation of the image by the level of the signals in the
whole spectral band was implemented here. (The user can choose between the normalisation
by global and local extrema). The contours show the levels of the narrowband power
in percents from the global spectral power maximum.

In the short period band (16\,--\,32 seconds) there is no clear localisation of the
perturbations, which occupy the whole magnetic fan and perhaps are
noise. Perturbations with longer periods are seen to be localised in
certain parts of the magnetic fan. The most pronounced
structuring is found for the spectral components in the range
128\,--\,256 seconds, which coincides with the result of periodmapping.
However, the narrowband maps obtained with PWF are much sharper and
cleaner. In particular, the narrowband maps show evidence of
splitting of the three-minute wave traces. Also, the wave traces are
stretched along the probable direction of the magnetic flux tubes.
Perturbations with the periods longer than 512 seconds that have
relatively low power, demonstrate some spatial structuring too.
Similar features have already been discussed in Sych and
Nakariakov (2008) and were shown to be standing waves.

%\begin{landscape}
%\end{landscape}

\subsection{Comparison of Wave Traces at Different Layers of the Solar Atmosphere}

Additional information about the waveguiding channels can be
obtained with the simultaneous use of EUV and the microwaves, \textit{e.g.}
the gyroresonant emission at the transition-region level observed
with NoRH at 17 GHz. A sequence of microwave polarisation (R--L)
images of AR 10756 on 4 May 2005, 02:30\,--\,05:00 UT was
synthesized with the temporal resolution ten seconds. At 17 GHz NoRH has the
spatial resolution of 10$^{\prime\prime}$. After standard preprocessing, the data
cube was analysed with the use of PWF. These results are compared
with the results of the PWF analysis of the TRACE EUV 171\AA\ data.

\begin{figure}    %%%%%%%%%%%%%%%%%% FIGURE 5
\resizebox{\hsize}{!}{\includegraphics[viewport=3 282 525 767]{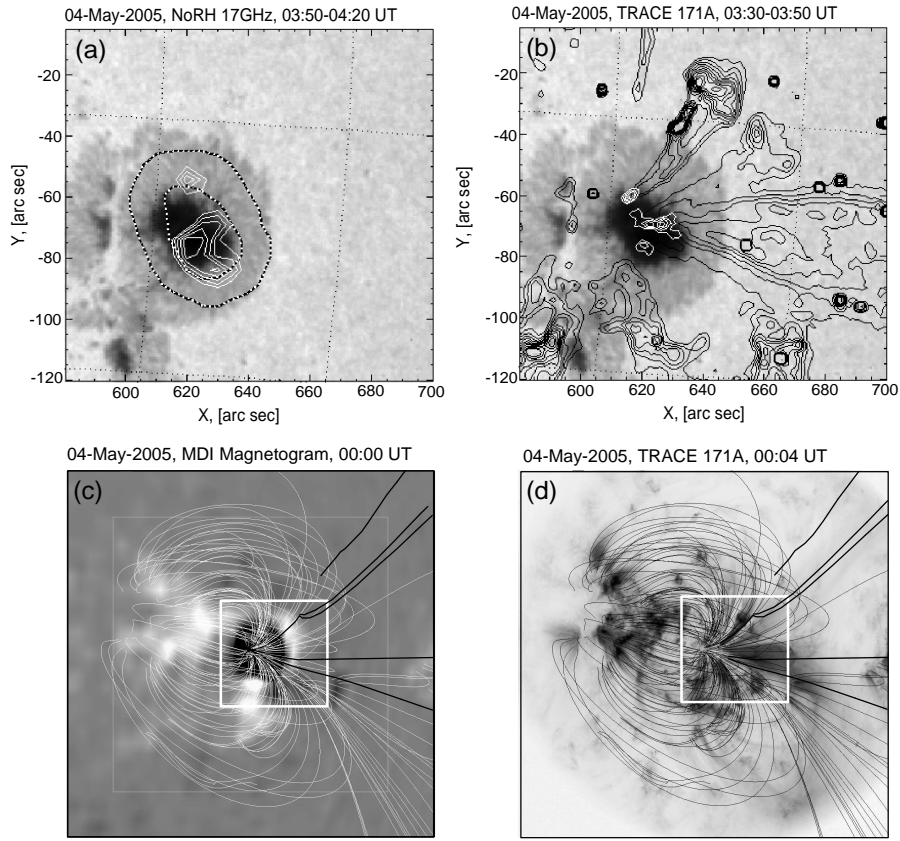}}
\caption{Panel (a) shows a white-light TRACE image of NOAA 10756 on 04 May 2005 with the three-minute wave traces (narrowband maps) obtained at 03:50\,--\,04:20 UT obtained with NoRH 17 GHz (white contours). The dotted black contours present the distribution of the 17 GHz microwave polarisation emission.   Panel (b) shows EUV narrowband three-minute wave traces in the corona at 03:30\,--\,03:50 UT (white contours). The black contours give the spatial structure of the broadband emission. Panels (c) and (d) shows the potential field interpolation of the coronal magnetic field overlapped at MDI magnetogram (00:00 UT) and TRACE 171\AA\ EUV emission (00:04 UT). The open magnetic field lines are shown by thick lines. The white box shows the position of the regions on panels (a) and (b).}
\label{fig5}
\end{figure}

Figure~\ref{fig5} shows narrowband maps of three-minute waves
obtained in the microwave and EUV bands, in panels (a) and (b)
respectively. The black contours give the spatial distribution of the
unfiltered microwave and EUV emission. The microwave polarisation
source has a symmetric shape and coincides with the sunspot umbra,
while the EUV emission sources are highly extended, highlighting
coronal loops. The narrowband map of the three-minute variations of the
microwave emission shows that there are two waves patterns of the size
about 20\,--\,30$^{\prime\prime}$. One located at (620, -75) in the umbra is
V-shaped. The other one at (620, -55), situated at the
umbra--penumbra boundary, has a round shape. The extended V-shaped
wave traces indicate the presence of certain waveguides, which are
very likely to be magnetic flux tubes (Van Doorsselaere \textit{et al.}, 2008;
see also discussion in Sych \textit{et al.}, 2009). Magnetic structuring above
the sunspot determines the detected structuring of the wave traces.
Thus, the narrowband-wave patterns reveal the geometry of the
waveguides which are the communication channels linking different
levels of the solar atmosphere.

Similar structuring is observed in the corona in the EUV band (Figure~\ref{fig5}b). The EUV data show that the three-minute wave traces are situated at the footpoints of coronal loops. Accounting for the difference in the spatial resolution of TRACE and NoRH, we conclude that three-minute wave traces of the EUV and microwave emission variations practically coincide. Thus, it is very likely that the variations are caused by the same wave process.

The geometry of the waveguiding channels was confirmed by the potential field extrapolation of the coronal magnetic field, obtained with \url{http://www.lmsal.com/forecast/TRACEview/}. Figure~\ref{fig5}c,d shows the magnetic geometry calculated at 00:00 and 00:04 UT, respectively. Interestingly, three-minute wave patterns are extended along the open magnetic-field lines. Perhaps, this suggests that the necessary condition for wave leakage to the corona is that the waveguiding magnetic flux tubes be open, or correspond to sufficiently long loops.

\section{Test Data (Temporal Approach)}

In the previous section we considered the spatial distribution of the narrowband variations that corresponded to a certain band of periods during a certain prescribed temporal interval. The temporal information about the evolution of the spectrum, which is also present in the 4D cube of the PWF analysis, is missing. However, this information can provide us with valuable knowledge of the amplitude, power, and phase dynamics of the oscillations. This is especially relevant to the corona, where the oscillations are usually rather non-stationary.

\subsection{Dynamics of the Three-Minute Radio Sources at the Transition Zone}
\label{sec61}

\begin{figure}    %%%%%%%%%%%%%%%%%% FIGURE 6
\resizebox{\hsize}{!}{\includegraphics[viewport=7 317 548 760]{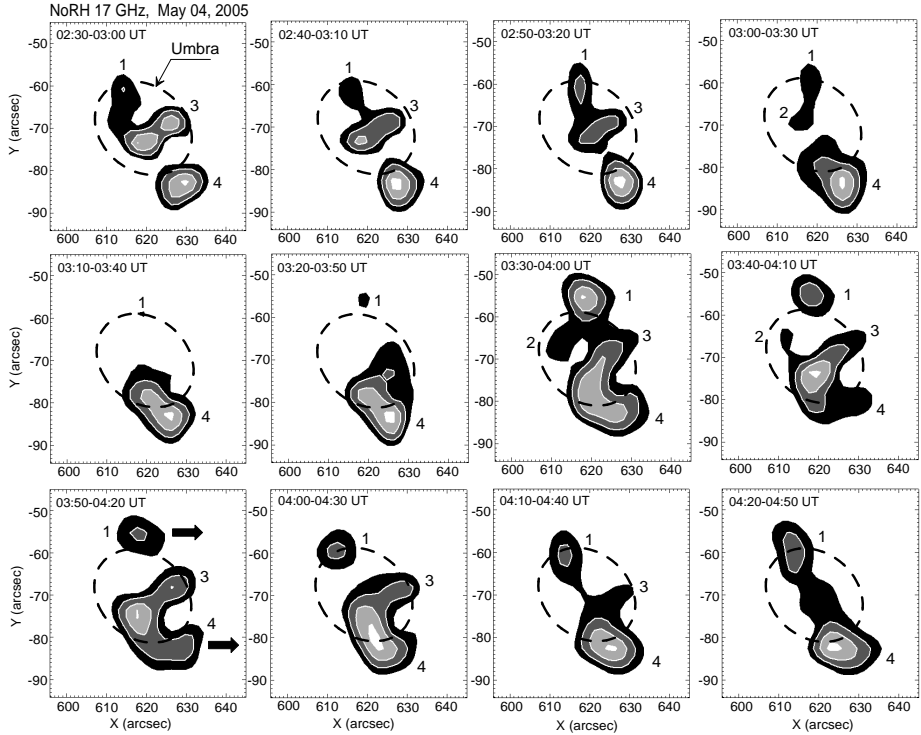}}
\caption{Temporal evolution of the narrowband three-minute wave traces of the microwave polarisation variation over the sunspot NOAA 10756, obtained with NoRH at 17 GHz on 04 May 2005, 02:30\,--\,04:50 UT.  The power of the oscillation is coded by the gray colour scale. The dashed curve indicates the umbra--penumbra boundary. The numbers label the wave traces. The arrows show the directions of the slits for the construction of the time--distance plots shown in Figure~\ref{fig7}. }
\label{fig6}
\end{figure}

The dynamics of three-minute waves in the transition region is studied in a sequence of narrowband maps constructed by the described system with the use of the harmonic PWF technique. The input datacube was formed by 17 GHz polarisation images obtained with NoRH on 04 May 2005 at 02:30\,--\,05:00 UT. The input for the creation of narrowband maps was a 30-minute time interval. Sliding the temporal interval by steps of ten minutes, we obtained twelve narrowband maps of the three-minute wave patterns, shown in Figure~\ref{fig6}. All maps are normalised. The dashed curve shows the sunspot-umbra boundary. The wave traces are numbered.

At the beginning of the observation (02:30\,--\,03:20 UT) there are three wave traces. Traces 1 and 4 are situated at the umbra--penumbra boundary, while trace 3 is in the umbra. During the initial stage of the observations, the shape remains almost constant. Later, since 03:00\,--\,03:30 UT traces 3 and 4 are changed, a new trace appears and a V-shaped structure is formed. Simultaneously, the strength of source 1 is growing. The maximum modification of the morphology is at 03:30\,--\,04:00 UT. Later on, the V-shaped structure disappears, and the initial configuration is restored.

The obtained changes in the morphology of the three-minute wave traces can be interpreted as the non-stationarity of the wave leakage from the sub-photospheric regions to the transition region. The wave traces are found to have round and extended shapes. This indicates the change of the wave vector component in the plane perpendicular to the LOS. In the initial stage of the observation, the wave traces had a round shape, indicating that the waves were propagating mainly along the LOS or were standing in the LOS direction. As the LOS was almost perpendicular to the solar surface in the vicinity of the sunspot, this corresponded to the vertical motions of plasma. Thus, the temporal evolution of the wave traces may indicate the change of the wave propagation direction.

\begin{figure}    %%%%%%%%%%%%%%%%%% FIGURE 7
\resizebox{\hsize}{!}{\includegraphics[viewport=0 243 554 780]{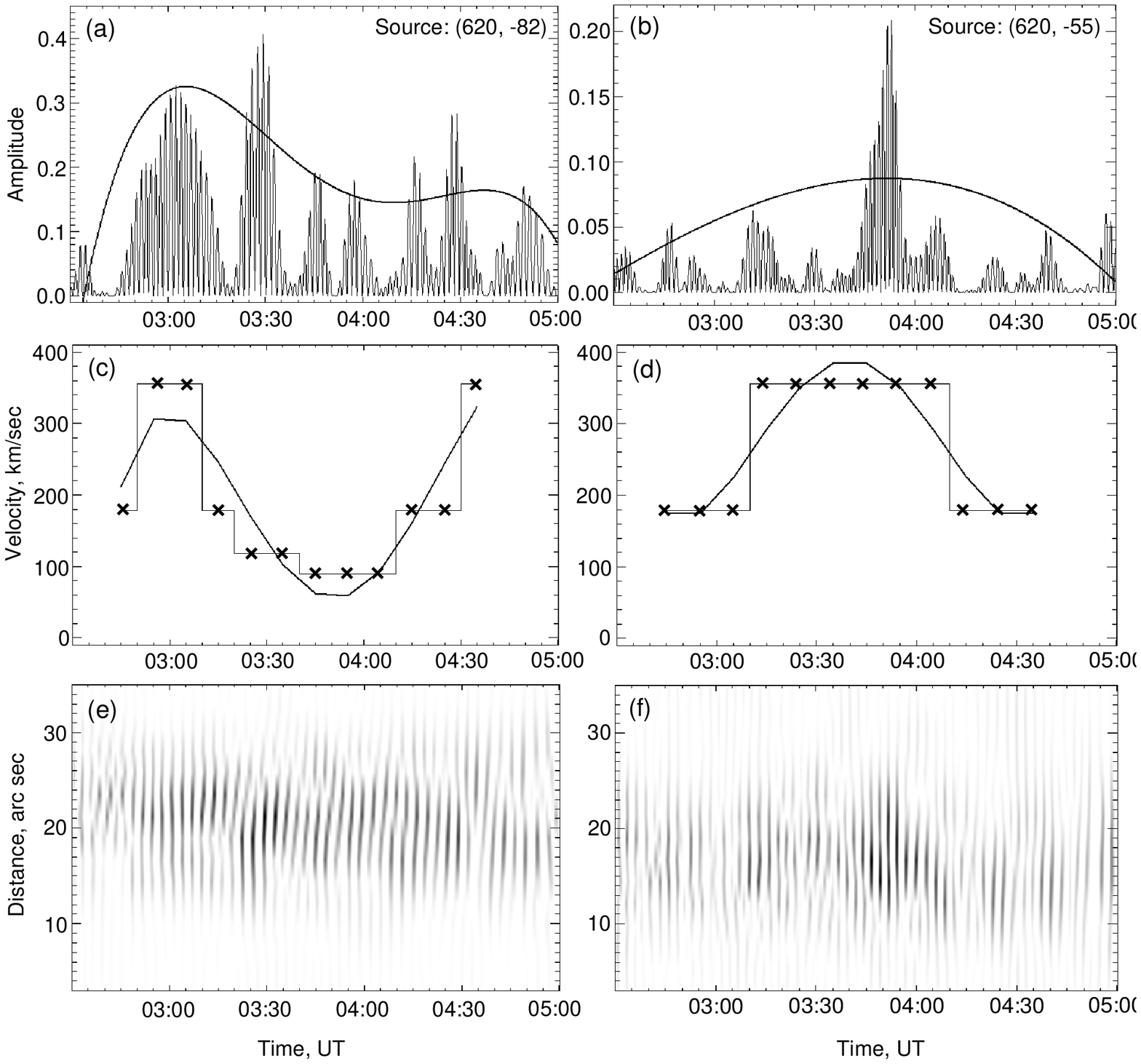}}
\caption{Temporal variation of three-minute waves at wave traces 1 (620, -82) (left column) and 4 (620, -55) (right column) along the horizontal slits shown in Figure~\ref{fig6} (03:50\,--\,04:20 UT). The upper panels show temporal variation of absolute amplitude of the filtered three-minute spectral component and its envelope (solid line). The middle panels show the time variation of the maximum projected phase speed and its envelope (solid line). The bottom panels show the time distance plots. }
\label{fig7}
\end{figure}

We would like to point out that such an orientation of the fingers of the
V-shaped structures in the three-minute narrowband maps of the microwave emission remained the
same throughout the passage of NOAA 10756 over the solar disk.
This indicates a continuous leakage of waves, which lasts for several days.
However, during energy releases, there can be relatively short temporal intervals (10\,--\,30 minutes)
when the orientation of the wave traces changes to the different from that of the quiet Sun
(Sych \textit{et al.}, 2009). After the energy releases, the configuration of the wave traces often returns back
to the pre-flare one.

It is interesting whether there is a relationship between the wave amplitude and the wave-propagation direction. To answer this question, we constructed time--distance plots of wave traces 1 and 4 in the time interval 02:30\,--\,05:00 UT along the slits directed along the arrows shown in Figure 6. The time-distance plots and their analysis are shown in Figure 7. The three-minute oscillations are detected during the whole observation period as almost vertical and oblique chevron structures in the time-distance plane. The gradient of the chevron structures allows one to estimate the phase speed of the observed waves, projected on the plane perpendicular to the LOS. The envelopes of the three-minute wave amplitudes and of the projected phase speeds show their high time correlation. Taking that the LOS is almost vertical, we obtain that the higher amplitude corresponds to the higher horizontal phase speed.

In the corona, phase speeds of propagating longitudinal waves are known to be systematically lower than the sound speed which is about 140\,--\,180 km s$^{-1}$ in the EUV band (\textit{e.g.} De Moortel, 2006; De Moortel, 2009). However in our observations the phase speed of the longitudinal waves in the transition region is found
to be 180\,--\,360 km s$^{-1}$, which is much higher than the transition region sound speed. The reason for this discrepancy is discussed in Conclusions.

\subsection{Dynamics of the Three-Minute EUV Sources in the Corona}

\begin{figure}    %%%%%%%%%%%%%%%%%% FIGURE 8
\resizebox{\hsize}{!}{\includegraphics[viewport=9 396 548 786]{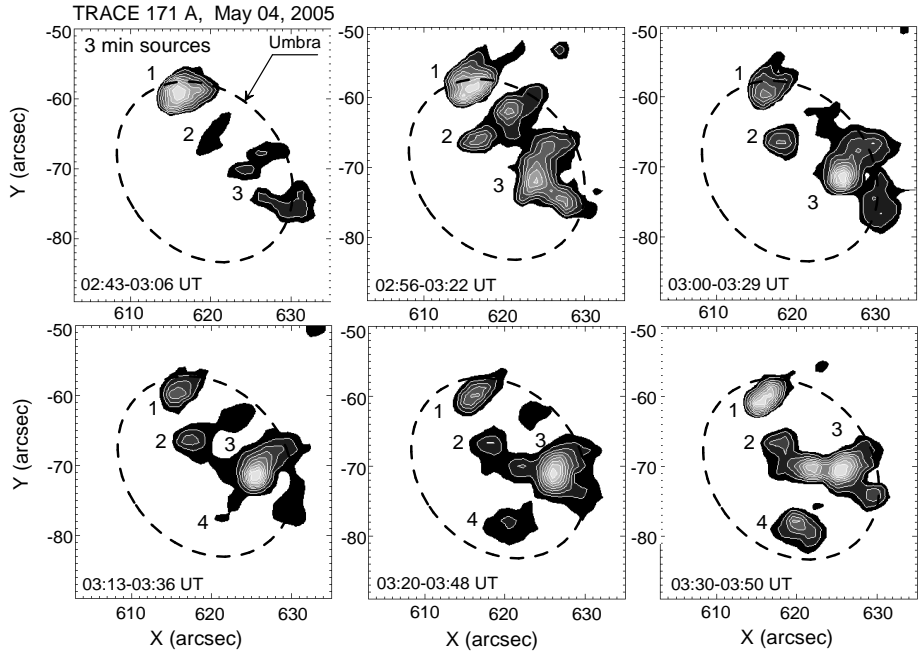}}
\caption{Time variation of the narrowband three-minute wave traces of the EUV emission variation over the sunspot NOAA 10756 obtained by TRACE 171\AA\ on 04 May 2005.  The power of the oscillation is coded by the gray colour scale. The dashed curve indicates the umbra--penumbra boundary. The numbers label the wave traces.} \label{fig8}
\end{figure}

For comparison, we considered the simultaneous observations of the same active region by TRACE in 171\AA\ bandpass, at 02:43\,--\,03:50 UT. The narrowband maps of the three-minute periodic variations of the EUV intensity were taken at 02:43\,--\,03:06, 02:56\,--\,03:22, 03:00\,--\,03:29, 03:13\,--\,03:36, 03:20\,--\,03:48 and 03:30\,--\,03:50 UT, shown in Figure~\ref{fig8}.

Similarly to the transition region level (see Figure~\ref{fig6}), the three-minute wave traces detected in the EUV band have fine structure. The enumeration of the wave traces coincides with the enumeration in Figure~\ref{fig6}. We would like to point out that the spatial resolution of TRACE is about an order of magnitude better than of NoRH, and hence the level of detail in the EUV narrowband maps is much higher than in microwaves. However, the application of the PWF technique highlights the wave motion of specific spatial details, which reduces the discrepancy.

The fine structure of the three-minute wave traces at the transition region and coronal levels is found to almost coincide. Thus, it is likely that these phenomena are caused by the same wave motion that propagates from the sunspot to the corona. For example, in the transition region, trace 1 (Figure~\ref{fig6}) coincides (accounting for the difference
in spatial resolution) with the coronal wave trace seen in Figure~\ref{fig8}. The magnetic field extrapolation (Figure~\ref{fig5}d) shows that the wave traces are situated on the open magnetic flux tubes. According to Section~\ref{sec61}, this trace correspond to a standing wave in the plane-of-the-sky. Thus, the observed waves are likely to be channelled in the vertical direction by the open magnetic flux tube from the transition region to the corona. The increase in the wave amplitude can cause the deformation of the wave trace pattern, see, \textit{e.g.}, the appearance of the V-shaped structures number 3 and 4 during 03:30\,--\,03:50~UT (Figure~\ref{fig8}).

As in the transition region, the projected phase speed measured with the use of the time--distance plot is found to be 
about 50 km s$^{-1}$. This value is consistent with the typical value for coronal propagating compressible waves 
(King \textit{et al.}, 2003; McEwan and De Moortel, 2006). \textbf{We should point out that in some works
the coronal propagating compressible waves, also known as longitudinal and slow magnetoacoustic (see, e.g.
Wang \textit{et al.}, 2009 and Marsh \textit{et al.}, 2009 for recent spectroscopic and stereoscopic results, 
respectively), are called \lq\lq periodic coronal outflows", see, \textit{e.g.} Sakao \textit{et al.}(2007).}

\section{Conclusions}

The principal aim of this paper is presentation of the specialised
computational resource \url{http://pwf.iszf.irk.ru} for the
automated detection and analysis of wave and oscillatory processes
in the solar atmosphere by remote access, and the demonstration of
its efficiency. The computational tool presented has a modular
structure, which includes algorithms for the automated localisation
of oscillatory processes in time and space, analysis of their
properties, and visualization of the results. The system allows for
easy change of the modules and parallel execution, which is crucial for
fast performance and for possible future applications for
real-time data analysis. The system is based upon the periodmapping
technique for the preliminary analysis and identification of the
region and times of interest, and the PWF technique for the detailed
analysis. The system provides for the determination of
the spatial-time-frequency-amplitude-phase dynamics and structure of
the waves. In particular, the system allows distinguishing between
standing (with respect to the LOS) and propagating waves. From the
user's point of view, the system is a web site that can be
interacted with through a standard Internet browser and does not
require installation of any specific software. We are not aware of
the existence of analogous resources in solar physics, or in other areas of science and technology.

The efficiency of the discussed system was tested by the analysis of
the three-minute waves over the sunspot NOAA 10756 observed in the EUV and
microwave bands. This study continues our previous work (Sych \textit{et.
al.}, 2009) on the phenomenological relationship between the
oscillatory processes in sunspot atmospheres and flaring energy
releases, such as flares and spikes. The main emphasis of the
present study was put on the analysis of three-minute waves in the
transition region and the corona in quite conditions. In
agreement with previous studies of this phenomenon (\textit{e.g.}, Berghmans
and Clette, 1999; De Moortel \textit{et al.}, 2000; De Moortel \textit{et al.}, 2002a,b;
Robbrecht \textit{et al.}, 2001; King \textit{et al.}, 2003; De Moortel, 2006), it was
found that the power of narrowband three-minute periodic variations of the
EUV and microwave emission was localised along the locally open
magnetic-field lines (determined by potential-field
extrapolation) near their footpoints. The regions of enhanced
power of the narrowband periodic variations, called wave traces,
highlight the waveguiding magnetised plasma structure. The spatial
extent of the wave traces along the magnetic field lines was found
to be about 20\,--\,30$^{\prime\prime}$, again in agreement with previous studies.

The long and short-period oscillations were found to be localised in
different regions over the sunspot. A similar feature has been
noticed \textit{e.g.} by Zhugzhda \textit{et al.} (2000) and Kobanov \textit{et al.} (2006) at
the photospheric and chromospheric levels, where three- and five-minute
oscillations were detected at different parts of a sunspot. In the
corona, this was established by De Moortel \textit{et al.} (2002), and Sych
and Nakariakov (2008). The three-minute wave traces determined at the
transition region and coronal levels are situated at the same
locations and have similar shape (taking into account an order-of-magnitude difference in the resolution). It is very likely that we
observe the same wave motion of the longitudinal mode, guided along
the magnetic field.

The analysis of the temporal evolution of three-minute wave traces revealed its
non-stationary nature. The projected phase speed was found to
correlate with the wave amplitude. The change of the projected wave
speed is likely to be attributed to the change in the magnetic-field
geometry, perhaps caused by the waves. In the transition region, the
projected phase speed is very high, 180\,--\,360 km$^{-1}$, which is much
higher than the sound speed at this height. We explain this
discrepancy as a projection effect. Indeed, when the longitudinal
three-minute waves propagate along the magnetic flux tubes in the corona,
they are localised in the transverse direction, and their transverse
wavelength is much smaller than the longitudinal wavelength. In
other words, the waves are essentially non-planar. This leads to the
apparent decrease in the observed projected phase speed with the
decrease in the angle between the LOS and the magnetic flux tube,
reported in a number of papers (\textit{e.g.} King \textit{et al.}, 2003).
However, below the corona, the decrease in the sound speed causes
the decrease in the longitudinal wavelength. Hence, the three-minute waves
become more planar. This leads to the increase in their projected
phase speed with the decrease in the angle between the LOS and the
direction of their propagation. (We would like to recall that the
projected phase speed of a plane wave propagating along the LOS is
infinite). Thus, we established that in the transition region the
growth of the three-minute amplitude is accompanied by the decrease in the
LOS angle to the wave-propagation direction. In other words, the
higher amplitude waves are observed to propagate more vertically. A
possible interpretation of this phenomenon is the effect of the
centrifugal force produced by the field-aligned flows of plasma
along the curved magnetic-field lines. This interpretation is
qualitatively consistent with the interpretation of the longitudinal
compressible coronal waves as slow magnetoacoustic waves (\textit{e.g.}
Nakariakov \textit{et al.}, 2000; De Moortel, 2006; De Moortel, 2009).

These findings demonstrate the efficiency of the described data analysis system, its ability to bring new interesting
scientific results, and its high potential for coronal seismology.

\acknowledgements  The work was supported by the Royal Society
UK-Russian International Joint Project, the grants RFBR
10-02-00153a, 08-02-13633-ofi-c, 08-02-91860-KO-a and 08-02-92204-GFEN-a. LO was
supported by NASA and NRL grants to Catholic University of America.
The TRACE and NoRH data were obtained from the TRACE and NoRH
database. The authors acknowledge the TRACE and NoRH consortia for
operating the instruments and performing the basic data reduction,
and especially  for the open data policy. Wavelet software was
provided by C. Torrence and G. Compo and is available at URL:
\url{http://paos.colorado.edu/research/wavelets/}

\end{article}
\end{document}